\documentclass[preprint,12pt]{elsarticle}

\usepackage{amssymb}
\usepackage{amsmath}

\journal{Sensors and Actuators A: Physical}

\graphicspath{ {./fig/} }

\usepackage{textgreek} 
\usepackage{soul,color} 

\begin{document}

\begin{frontmatter}



\title{Laser fabrication of Ti stent and facile MEMS flow sensor integration for implantable respiration monitoring}

\author[hcu]{Muhammad Salman Al Farisi\corref{cor1}}
\ead{alfarisi@hiroshima-cu.ac.jp}
\author[hcu]{Takuya Kawata}
\author[hcu]{Yoshihiro Hasegawa}
\author[hcu]{Mohammad Nizar Mohamed Zukri}
\author[nagoya]{Miyoko Matsushima}
\author[nagoya]{Tsutomu Kawabe}
\author[hcu]{Mitsuhiro Shikida}
\cortext[cor1]{Corresponding author}

\affiliation[hcu]{organization={Department of Biomedical Information Sciences,
								Hiroshima City University},
            city={Hiroshima},
            postcode={731-3194},
            country={Japan}}

\affiliation[nagoya]{organization={Division of Host Defense Sciences, Omics Health Sciences,
									Department of Integrated Health Sciences, 
									Graduate School of Medicine, Nagoya University},
            city={Nagoya},
            postcode={461-8673},
            country={Japan}}



\begin{abstract}
Animal experiments play a vital role in drug discovery and development by providing essential data on a drug's efficacy, safety, and physiological effects before advancing to human clinical trials. In this study, we propose a stent-based flow sensor designed to measure airflow in the airways of laboratory animals. The stent was fabricated from biocompatible Ti using a combination of fiber laser digital processing and an origami-inspired folding technique. The sensing structure was developed through standard micro-electromechanical systems (MEMS) microfabrication technology. To integrate the sensing structure with the metallic stent, a facile insertion process was employed, where the sensor film was positioned at the stent's center using its natural buckling mechanism. Once fabricated, the stent implant was expanded and installed within an airway-mimicking tube to validate its functionality. A proof-of-concept trial using an artificial ventilator successfully demonstrated real-time respiration monitoring, confirming the feasibility of the proposed system for airflow measurement in preclinical studies. This stent-based sensor offers a promising approach for enhancing respiratory assessments in laboratory animals, potentially improving the accuracy of drug evaluations and respiratory disease research.
\end{abstract}



\begin{keyword}
Respiration \sep stent \sep airflow sensor \sep implant \sep animal experiment


\end{keyword}

\end{frontmatter}



\section{Introduction}\label{sec:intro}

Animal experiments are an essential component of drug discovery and development, providing crucial data on a drug’s efficacy, safety, and physiological effects before advancing to human clinical trials~\cite{Swindle2012,Bryda2013}. These studies allow researchers to assess how a drug interacts with living organisms, including its pharmacokinetics, i.e. how it is absorbed, distributed, metabolized, and excreted, and its pharmacodynamics, which describes the biological mechanisms and effects of the drug~\cite{Mukherjee2022}. Unlike in vitro studies, which provide limited insights into systemic interactions, animal models enable scientists to observe how drugs behave in complex biological systems that closely resemble human physiology~\cite{Esteves2018}. By using disease-specific animal models, researchers can evaluate the therapeutic potential of drug candidates while also identifying possible toxicities and adverse effects that could pose risks in human trials~\cite{Singh2021}. To ensure a sufficient safety profile, regulatory agencies such as the U.S. Food and Drug Administration (FDA), the European Medicines Agency (EMA), and Japan’s Pharmaceuticals and Medical Devices Agency (PMDA) mandate preclinical animal testing before human clinical trials can commence~\cite{Tanimoto2013}.  

Besides cardiovascular monitoring, respiration monitoring in experimental animals is also essential for understanding respiratory physiology, assessing drug effects, and evaluating how the disease progresses. High mortality rate caused by respiratory-related diseases further emphasized the importance of respiration monitoring during drug development~\cite{Mathers2006,Foreman2018,Crook2021}. Several methods are employed to track respiration, each with varying degrees of invasiveness and accuracy~\cite{Grimaud2018}. One approach is measuring respiratory muscle activity using electromyography (EMG) of the diaphragm or other inspiratory muscles~\cite{Li2016}. EMG provides detailed information about respiratory cycles and volumes but requires surgical implantation of electrodes, making it highly invasive. Needle EMG has also emerged as a non-invasive alternative of traditional EMG, in which careful placement and operation was necessary to reduce tissue injury induced by the needle~\cite{Podnar2016}. Additionally, nonrespiratory muscles like the genioglossus have been explored~\cite{Cui2016}, though their reliability for respiratory monitoring remains uncertain. 

Another category of respiration monitoring focuses on air movement. This traditional approach can be traced back to conventional spirometry performed for human medical check ups~\cite{Johns2014,Ohkura2021}. Intubation, where a cannula is inserted into the trachea of an anesthetized animal, allows for precise airflow and volume measurements~\cite{AlFarisi2023,Nagayama2024}. Similarly, intranasal cannulas and face masks can be used to track breathing in a less invasive manner~\cite{Bolding2017}. However, fixing such mask on unrestrained animals have been a huge challenge. Lung plethysmography, which includes whole-body and airflow-based variations, provides real-time measurements of breathing rate and tidal volume~\cite{Lim2014}. It enabled measurement of respiratory rate and tidal volume by detecting pressure changes within an enclosed chamber. However, its theoretical model has been under scrutiny for accuracy in respiratory monitoring~\cite{Bates2004,Lundblad2007}.


Non-contact methods are also gaining traction, particularly for monitoring freely moving animals. Body movement sensors, such as piezoelectric devices~\cite{Flores2007,Horie2024a}, detect respiratory-induced motion but may require the subject to remain stationary for accuracy. Infrared cameras and video monitoring offer a completely noninvasive alternative by tracking temperature changes around the nostrils or analyzing chest movements~\cite{EsquivelzetaRabell2017,Kurnikova2017}. These methods, while promising, often require sophisticated image processing and may not yet provide the same level of accuracy as direct airflow measurements.

Considering various measurement principle, we focused on direct airflow measurement which best reflect the respiration condition of experimental animals. In particular, micro-electro mechanical systems (MEMS)-based thermal flow sensors have emerged as a promising tool for simultaneously measuring both respiratory and cardiovascular parameters through airflow analysis in the respiratory tract~\cite{AlFarisi2023,Nagayama2024}. Previously, we have proposed a stent-integrated MEMS airflow sensor, demonstrating simultaneous measurement of the respiration and heartbeat~\cite{Noma2021}. However, conventional MEMS materials were used, which lacked biocompatibility. In our previous study, a Cu stent was fabricated through a standard metal etching microfabrication process. While Cu possesses excellent electrical and thermal conductivity, it is not widely used in implants because it can induce cytotoxicity and trigger adverse biological responses when in prolonged contact with body tissues. In addition, the fabrication process was complicated, involving multiple microfabrication process steps and sensor tube integration using multiple resin tubes. This has hindered the system from practical application.

As a solution, here we proposed a facile fabrication process to produce a biocompatible stent implant sensor system. Ti was selected as the stent material. Cu and Ti differ significantly in their suitability for biomedical implants due to their distinct material properties and biocompatibility. Ti is renowned for its exceptional biocompatibility, corrosion resistance, and strength-to-weight ratio, making it a preferred material for a wide range of biomedical implants, including dental, orthopedic, and cardiovascular devices. The long term stability of the proposed system can be expected taking advantage from the corrosion rate of Ti implants of several \textmu m/year~\cite{Pound2014}, and the chemical inertness of the Au electrode utilized as the sensing structure. The stent fabrication involved a single step laser processing to produce kirigami planar structure and origami technique to form it into tubular stent structure. In addition, a facile integration process of a MEMS airflow sensing structure is also proposed, which could bring the system closer to clinical implementation.



\section{Experimental Procedure}

\subsection{MEMS airflow sensor}

\begin{figure}[tb]
\centering
\includegraphics[width=0.5\columnwidth]{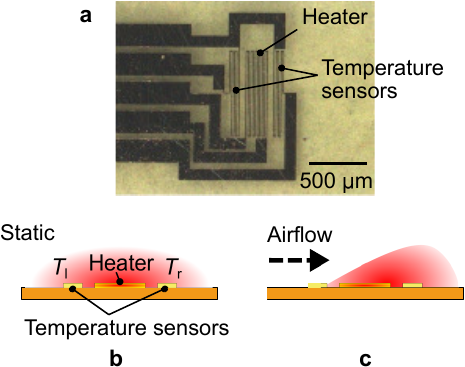}
\caption{Respiration airflow sensing using thermal calorimetry.
(a) Micrograph of the fabricated MEMS thermal flow sensing structure.
Sensing mechanism using heat distribution around the sensing structure in the
(a) absence and (b) presence of airflow.}
\label{fig:flow-mechanism}
\end{figure}

\begin{figure}[tb]
\centering
\includegraphics[width=0.45\columnwidth]{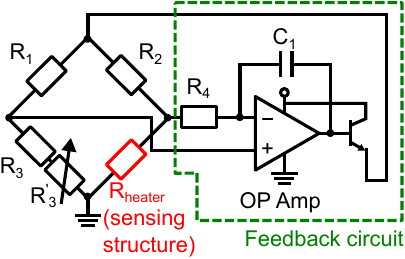}
\caption{Externally installed heater actuation circuit.}
\label{fig:circuit}
\end{figure}

The respiration measurement was performed using MEMS thermal flow sensor fabricated on a 7.5 \textmu m thin polyimide (PI) film substrate. The flow sensor film was fabricated using a standard microfabrication lift-off process as reported in our previous study~\cite{AlFarisi2023}. Au thin film of 250 nm thickness was utilized as the sensing structure metallization. The sensor worked based on the thermal calorimetry mechanism using serpentine thin-film metallic structures as the sensing element as shown in Fig.~\ref{fig:flow-mechanism} (a). The cross-section of the sensing structure is schematically depicted in Figs.~\ref{fig:flow-mechanism} (b) and (c). It consist of a heater and 2 temperature sensing structures, each at the heater's upstream and downstream direction. All these 3 structures worked based on the Joule heating of metallic thin-film wires. The heater was driven at a constant temperature using a feedback control circuit as shown in Fig.~\ref{fig:circuit}~\cite{AlFarisi2025}. The feedback control circuit was located separately from the sensor, connected by a flexible printed circuit wiring. Here $R_\mathrm{heater}$ indicates the heater structure in the sensor. In the electrical circuit, the Wheatstone bridge's balance was maintained as mathematically formulated in Eq.~\ref{eq:bridge}.

\begin{equation}
\label{eq:bridge}
R_1 R_\mathrm{heater} = R_2 (R_3 + R'_3)
\end{equation}

Therefore, by setting the magnitudes of fixed resistors $R_1$, $R_2$, $R_3$ and variable resistor $R'_3$, the resistance of the heater can be adjusted. According to the temperature coefficient of resistance (TCR) of the heater's material $\alpha$, the heater's resistance can be adjusted so that it is driven at a certain temperature. The resistance at the set temperature $R'_{\mathrm{heater}}$ is mathematically formulated as in Eq.~\ref{eq:tcr}.

\begin{equation}
\label{eq:tcr}
R'_\mathrm{heater} = R_\mathrm{heater} \{1 + \alpha (T - T_0)\}
\end{equation}

Here, $T$ indicates the target driving temperature, and $T_0$ indicates the room temperature where the heater's initial resistance $R_\mathrm{heater}$ was measured. The heating temperature was designed at a higher temperature than the room or body temperature to allow flow measurement, while also taking into account the biomedical compatibility. The sensitivity of thermal sensors usually rise with the heaters' driving temperature~\cite{AlFarisi2023b}, and therefore it should be maximized. Meanwhile, animal body temperature can be around 37$^\circ$C, and significant increase from this temperature may cause harm to the body, although most of the heat is dissipated at less than a few mm distance from the heater.For instance, less than 20\% of the overtemperature remains at 0.5 mm and less than 10\% remains at 1 mm distance from the heater~\cite{AlFarisi2023}. Therefore, the heater was driven at 50$^\circ$C in this study.

After driving the heater at the above temperature, the electrical circuit worked to maintain the heater at constant temperature through the Wheatstone bridge. When the airflow inside the respiratory airway is not flowing, the heater maintains the set temperature. And when the respiration airflow is flowing, the heat is dissipated from the heater, which causes the heater to lose its temperature and resistance momentarily. Such a resistance drop disturbed the bridge's balance and the potential difference is immediately transferred to the op-amp, which in turn supply additional voltage to the bridge to maintain its balance. This additional voltage results in additional current flowing through the heater, which maintains the heater constantly at the set temperature.

When the airflow is absent, the heat dissipated from the heater is distributed evenly. Therefore, the readings of both the temperature sensors at its upstream and downstream directions are in balance. When there present an airflow, the heat is dissipated from the heater from the upstream to the downstream direction. This disturbed the balance of the temperature sensors' readings. The differential of the temperature sensors' readings corresponds to the flow rate, which is the thermal calorimetric flow rate sensing mechanism employed in this study. The resistance change of both sensors was measured using a differential amplifier circuit.

The performance of such MEMS thermal flow sensor also relies on the thermal mass around the sensing structure~\cite{AlFarisi2023}. Therefore, in this study, a thin PI film was utilized as the sensor substrate to limit the thermal mass by the substrate. The sensor film was attached to a polyethylene terephthalate (PET) film with 50~\textmu m thickness to introduce structural rigidity in a certain degree, allowing its seamless installation within the stent. A cavity was formed on the PET support film underneath the sensing structure to limit the thermal mass.




\subsection{Laser fabrication of Ti stent}

\begin{table}
\renewcommand\arraystretch{1.5}
\centering
\caption{Fiber laser optimized processing parameters.}
\label{table:lasercut}
\footnotesize
\vspace*{2mm}
\begin{tabular}{p{30mm} c}
\hline
Parameter & Set value \\
\hline
Laser power	& 14 W 	\\
Scan speed	& 5 mm/s		\\
Pulse width		& 250 ns	\\
Pulse frequency	& 5 kHz	\\
\hline
\end{tabular}
\end{table}

\begin{figure}[tbp]
\centering
\includegraphics[width=0.7\columnwidth]{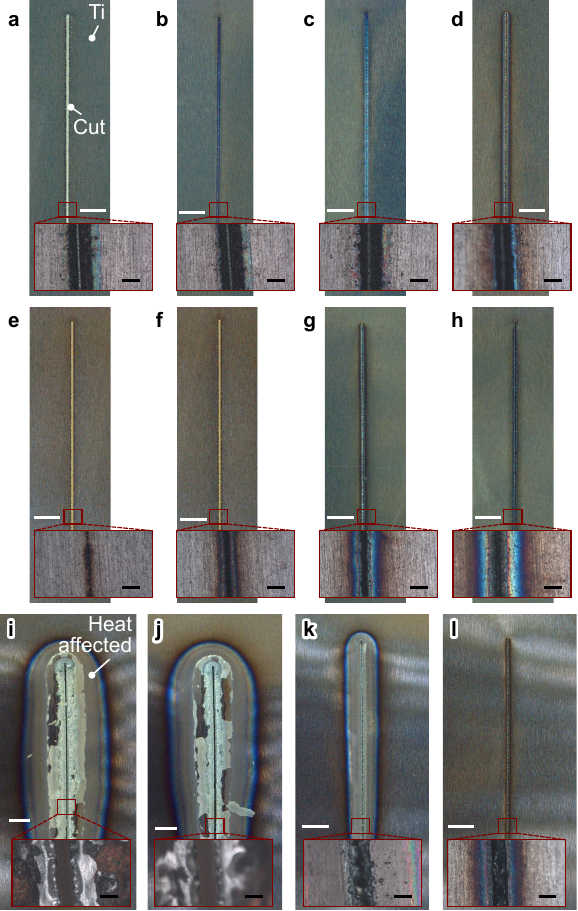}
\caption{Laser ablated Ti foil with variable laser power of
(a) 8 W, (b) 10 W, (c) 12 W, and (d) 14 W;
variable laser scanning speed of
(e) 50 mm/s, (f) 20 mm/s, (g) 10 mm/s, and (h) 5 mm/s;
variable laser emission frequency of
(i) 50 kHz, (j) 25 kHz, (k) 10 kHz, and (l) 5 kHz.
Scale bars are of 1 mm length.
Scale bars of the inset micrographs are of 10 \textmu m length.
}
\label{fig:lasercut}
\end{figure}

\begin{figure}[tb]
\centering
\includegraphics[width=0.4\columnwidth]{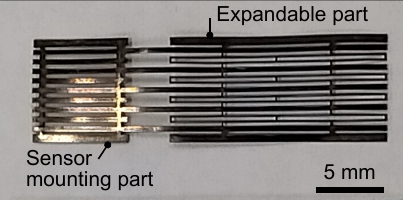}
\caption{Planar Ti stent structure.}
\label{fig:stent}
\end{figure}

The Ti stent was fabricated through a kirigami-origami process utilizing a Ti foil with 50 \textmu m thickness patterned using a fiber laser (LM110M, SmartDIYs Co., Ltd.). Fiber laser patterning involves thermal energy to ablate the metallic foil. Such energy is reflected by the laser power and laser scanning speed. By introducing a nanosecond pulse processing, the laser ablation process can be alternated with rests, which induced cooling steps allowing patterning with low damage. The laser processing condition was optimized considering the required ablation energy.

For this optimization, straight lines of 10 mm in length was patterned. Three primary parameters: laser power, speed, and frequency, were studied and adjusted to optimize the fine laser cutting of 50~\textmu m thick Ti foil. After conducting the experiments, these parameters were finalized to serve as a reference recipe for cutting Ti foil of this thickness. The pulse width was kept constant at 250 ns laser on duration, adhering to the default settings for metal cutting with the fiber laser. Table~\ref{table:lasercut} presents the optimized setting for the laser cutting parameters used for stent fabrication in this study. Variation of each the laser ablation power, scanning speed or pulse frequency resulted in imperfect cuts as explained below. The parameters other than the variable parameter in each experiment were kept constant in accordance to the optimum parameters as mentioned in Table~\ref{table:lasercut}.

Laser cutting results of 50~\textmu m Ti foil under several parameter are depicted in Fig.~\ref{fig:lasercut}. Fig.~\ref{fig:lasercut} (a) to (d) demonstrates the effects of laser cutting using a fiber laser at different laser power: 8 W, 10 W, 12 W, and 14 W; under the fixed scanning speed (5 mm/s), pulse width (250 ns) and frequency (5 kHz). The combination of both the laser power and the laser scanning speed determines the ablation energy. Fig.~\ref{fig:lasercut} (a) patterned at 8 W shows insufficient energy absorption, indicating incomplete cuts. The laser power increase led to increase energy absorption. The cutting through was observed at 10 W and 12 W as in Figs.~\ref{fig:lasercut} (b) and (c). However the lines observed incomplete cuts. The optimal cutting condition was achieved at 14 W as in Fig.~\ref{fig:lasercut} (d), where a clean, smooth, and precise cutting line was observed. Large magnification micrograph inset of the cutting interface also indicates that the resulting surface was less rough in this condition in comparison to the others.


Fig.~\ref{fig:lasercut} (e) to (h) demonstrates the effects of laser cutting using the fiber laser at different laser scanning speed: 50 mm/s, 20 mm/s, 10 mm/s, and 5 mm/s; under the fixed laser power (14 W), pulse width (250 ns) and frequency (5 kHz). The laser cutting Fig~\ref{fig:lasercut} (e) and (f) shows that processing at 50 mm/s and 20 mm/s yielded insufficient energy absorption, indicating incomplete cuts. As the speed decreases, the prolonged exposure to heat lead to increase energy absorption. The cutting through was observed at 10 mm/s in Fig.~\ref{fig:lasercut} (g), however rougher edges are observed due to incomplete cuts. The optimal cutting condition was achieved at 5 mm/s Fig.~\ref{fig:lasercut} (h), where a clean cuts, smooth, and precise cutting line was observed. The inset micrograph also indicates that the resulting surface was less rough in this condition in comparison to the others.

Fig~\ref{fig:lasercut} (i) to (l) demonstrates the effects of laser cutting using a fiber laser at different frequencies: 50 kHz, 25 kHz, 10 kHz, and 5 kHz; under the fixed laser power (14 W), scanning speed (5 mm/s) and pulse width (250 ns). Fig.~\ref{fig:lasercut} (i) shows significant heat-affected zones with excessive debris, indicating overheating caused by the high frequency. The processing frequency reflects how often the resting/cooling step is introduced, while the laser on duration was maintained constant at 250 ns. As the frequency decreases, the cutting quality improved gradually, with reduced thermal effects and smoother edges observed at 25 kHz and 10 kHz in Fig.~\ref{fig:lasercut} (j) and (k). The optimal cutting condition was achieved at 5 kHz as shown in Fig.~\ref{fig:lasercut} (d), where a clear, smooth, and precise cutting line was observed with minimal debris and negligible heat-affected zones as also confirmed by the micrograph inset. Below this frequency, no cut was observed.

Using the optimized laser cutting parameters as in Table~\ref{table:lasercut}, a stent structure was designed and implemented with the fiber laser. One of the main advantages of the proposed laser processing is that it offers digital rapid prototyping by directly implementing the design. The optimized parameter yielded the least surface roughness and surface damage, which is crucial since the stent will be in contact with the airway wall. To further reduce the surface roughness, wet etching process can be introduced. By such an additional short time wet etching process, sharp tips resulted from the laser processing can be annihilated. The stent planar design consisted of a sensing mounting area and an expandable stent area. The fabricated Ti stent is depicted in Fig.~\ref{fig:stent}.

\subsection{Sensor integration}

\begin{figure*}[tb]
\centering
\includegraphics[width=\columnwidth]{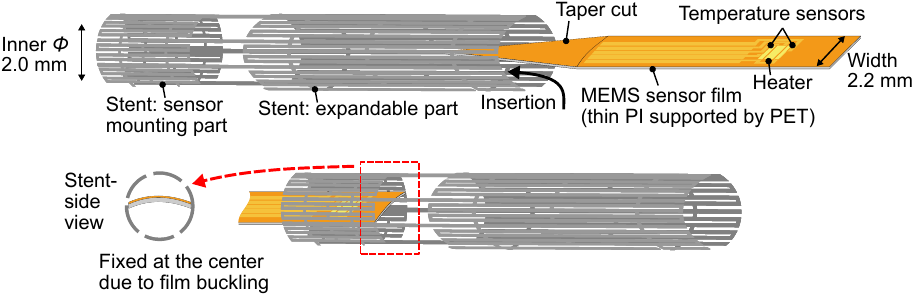}
\caption{Sensor integration through a facile insertion process.}
\label{fig:insertion}
\end{figure*}

\begin{figure}[tb]
\centering
\includegraphics[width=0.42\columnwidth]{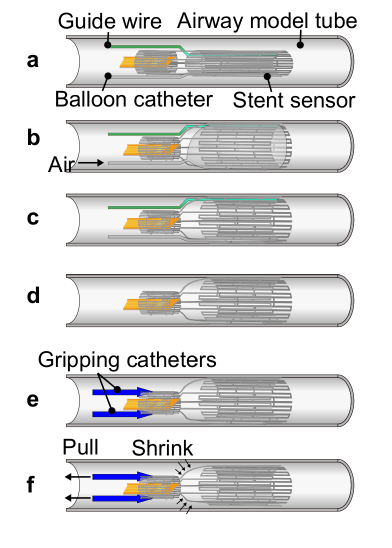}
\caption{Stent deployment and extraction mechanism.
(a) Stent insertion using a guide wire and a balloon catheter.
(b) Balloon inflation to expand the stent.
(c) Balloon deflation.
(d) Ballon catheter ejection.
(e) Gripping non-expanded part of the stent.
(f) Pull out the stent and the expanded part shrinks
by the pulling force and friction with the airway.
}
\label{fig:installation}
\end{figure}

\begin{figure}[tb]
\centering
\includegraphics[width=0.65\columnwidth]{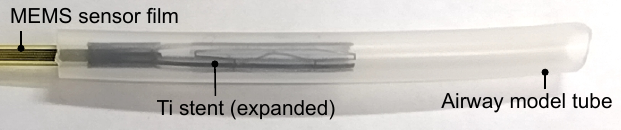}
\caption{The proposed device installed inside an airway model.}
\label{fig:stent-expanded}
\end{figure}

The fabricated planar stent was formed into a cylindrical structure through an origami technique by wrapping it around the outer surface of a perfluoroalkoxy alkane (PFA) tube, which served as a temporary core. Both ends of the stent were secured using adhesive to maintain its shape, and the PFA tube was subsequently removed. A tube with an external diameter of 2.0 mm was selected to ensure proper fitting in both the sensor mounting region and the expandable stent section. This approach allowed for seamless integration between the sensing component and the stent structure as illustrated in Fig.~\ref{fig:insertion}.

The sensor film was cut into a stick-like shape with a tapered end to facilitate its insertion into the sensor mounting area within the stent. The taper guided the film smoothly into position, while its gradually increasing width ensured it remained centered during the pulling insertion process. The maximum width of the polyimide (PI) film was 2.2 mm, slightly exceeding the stent’s internal diameter of 2.0 mm. This intentional size difference allowed the film to buckle upon proper placement, ensuring secure positioning at the center of the tube. Buckling occurs when a thin film experiences compressive stress from opposing directions, enabling the sensing structure to self-align within the tube, simplifying the installation process.

The stent was deployed into an airway mimicking silicone tube as illustrated in Fig.~\ref{fig:installation}. The stent was inserted into the airway with its expandable part attached to a deflated balloon catheter as illustrated in Fig.~\ref{fig:installation} (a). The balloon was then inflated as shown in Fig.~\ref{fig:installation} (b) to expand the stent expandable part and firmly fix the stent at the desired position. Then the balloon was deflated by releasing the supplied air and extracted from the airway as illustrated in Figs.~\ref{fig:installation} (c) and (d). Here the expandable part of the stent played a role to fix the stent within the airway, while the sensor mounting part housed the sensor film firmly. The integrity of film sensor installed inside a tubular housing utilizing its buckling have been tested under various airflow rate conditions, including an animal experiment in our previous study~\cite{AlFarisi2023}. During practical implementation, the stent extraction can be performed using one or several gripping catheters as illustrated in Fig.~\ref{fig:installation} (e). The gripper is used to catch the non-expanded sensor mounting part of the stent. By slight pulling, the pulling force and the friction with aiway wall causes the expanded part of the stent to shrink as illustrated in Fig.~\ref{fig:installation} (f). This way, the stent is released from the airway and the extraction can be performed by further pulling the gripping catheters.

Combined with the stent, the sensing structure is installed at around the center of the airway, ensuring measurement at the position where the airflow velocity is maximum. The fabricated stent device installed inside an airway mimicking silicone tube is depicted in Fig.~\ref{fig:stent-expanded}. Here the expandable part sticked to the inner wall of the airway, where the airflow velocity is minimum, which will minimize its impact to the ariflow dynamics. Meanwhile, the sensor mounting part and the sensor were installed at the center of the airway, occuppying around 5\% of the cross-sectional area. This ratio indicates the airway resistance induced by the proposed stent, which can be minimized by thinning the stent material or the sensor, at the cost of the system rigidity and structural stability.

\section{Results and Discussion}

\subsection{Temperature coefficient of resistance}

\begin{figure}[tb]
\centering
\includegraphics[width=0.45\columnwidth]{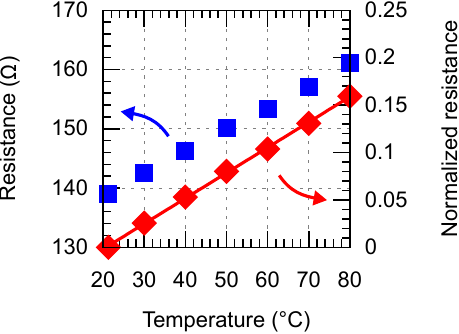}
\caption{TCR measurement of the sensing structure.}
\label{fig:tcr}
\end{figure}

First, the temperature coefficient of resistance (TCR) of the heater element in the film sensor was evaluated. This assessment is crucial for ensuring accurate heating temperature adjustment inside the body. Since excessive heat could degrade damage the body tissue, the flow sensor must operate at a low temperature. To allow such this, we established a calibration curve by determining the relationship between the temperature and the resistance of the heater. The heater was incorporated into a voltage divider circuit, where a small current was applied to prevent self-heating effects that could interfere with the measurement. The heater was then placed in a thermostatic oven, and its resistance was recorded as the temperature was gradually varied.

The results, shown in Fig.~\ref{fig:tcr}, demonstrate that the heater's resistance increased linearly from 138.9 $\Omega$ at room temperature of 21.3$^\circ$C to 161.1 $\Omega$ at 80$^\circ$C. To standardize these findings, the resistance change was normalized using the room temperature resistance, yielding a TCR value of 2,677 ppm/$^\circ$C. This value provided a basis for controlling the operating temperature of the flow sensor in subsequent experiments. By adjusting the driving temperature according to the obtained calibration curve, we ensured that the sensor maintained optimal performance at 50$^\circ$C heater actuation temperature, thereby preserving the safety during measurement. To further enhance the safety procedure, the sensing structure was also set at the middle of the airway, where a certain distance was provided to allow the heat to dissipate in the air before reaching the body tissue. 

\subsection{Fundamental airflow sensing characteristics}

\begin{figure}[tb]
\centering
\includegraphics[width=0.5\columnwidth]{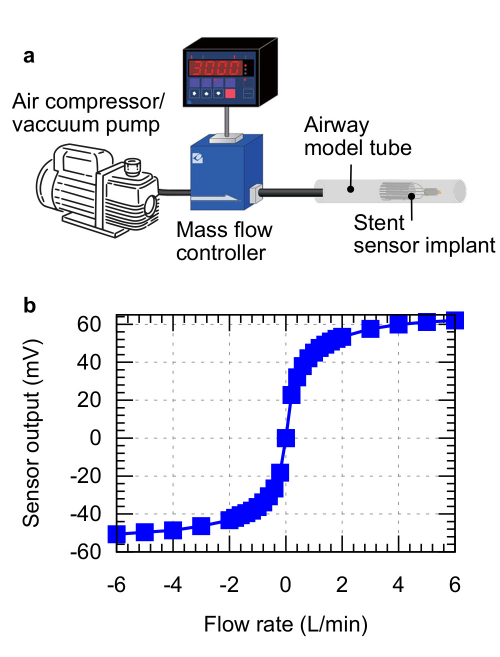}
\caption{Calibration experimental (a) setup and (b) result
of the sensor embedded in an airway model tube.}
\label{fig:calibration}
\end{figure}

The MEMS flow sensor integrated stent device was then installed inside a silicone rubber tube with an internal diameter of 3.0 mm, which mimics the airway of rabbits~\cite{Arendt2023}, one of the typical experimental animals~\cite{Keir2008}. Following the installation process, the fundamental characteristics of the fabricated sensor were evaluated using a precisely controlled airflow supply. The experimental setup for characterization is illustrated in Fig.~\ref{fig:calibration} (a). Airflow was generated either by an air compressor for positive flow or by a vacuum pump for negative flow. A commercially available mass flow controller (Model 3200, KOFLOC Corp.) regulated the airflow rate to ensure accuracy and consistency. To capture the sensor’s response, the temperature sensors were connected to a Wheatstone bridge circuit with a gain of 100, allowing the thermal output to be converted into a voltage signal for analysis.

The airflow sensing performance of each packaged sensor was then assessed under controlled conditions. The heater was operated at 50$^\circ$C using the constant temperature feedback-controlled circuit to maintain stable operation. The ambient temperature was 22.1$^\circ$C, which reflects the air temperature flowing through the sensor. Even though the temperature inside a living body is generally higher than this room temperature, there is minimal temperature variation in an individual (within around 1$^\circ$C). In addition, the temperature difference between the inhalation and exhalation is also limited because the airflow temperature in both conditions are heavily influenced by the body temperature. Therefore, during practical application the sensor can be calibrated using airflow at the expected body temperature of the target application with minimal modification to the proposed sensing structure. An additional temperature sensor can be embedded to the sensing structure to compensate the airflow temperature~\cite{AlFarisi2024}. The calorimetric sensor’s output voltage was recorded in response to varying airflow rates, as shown in Fig.~\ref{fig:calibration} (b). The resulting data was used to generate a calibration curve, which served as a reference for subsequent proof-of-concept experiments. This calibration process ensured that the sensor could provide reliable and accurate airflow measurements for future applications.

\subsection{Proof-of-concept experiment}

\begin{figure}[tb]
\centering
\includegraphics[width=0.45\columnwidth]{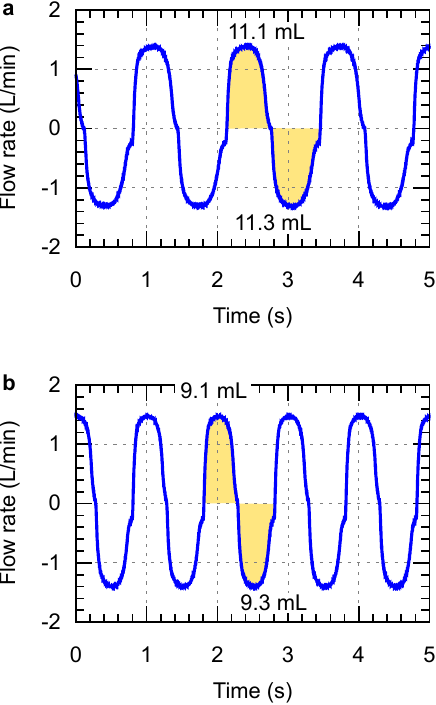}
\caption{Proof-of-concept experimental results under artificial ventilation operation at
(a) 45 and (b) 60 breaths/min.}
\label{fig:ventilator}
\end{figure}

To evaluate the sensing capability of the proposed sensor, its performance was tested in measuring reciprocating airflow generated by an artificial ventilator (Small Animal Ventilator Model 683, Harvard Apparatus) through the same silicone tube installation as the calibration experiment. The ventilator was configured to deliver airflow at a frequency of 45 and 60 breaths/min with an approximate volume of 10 mL per cycle, resembling the typical respiration of rabbits~\cite{Johnson-Delaney2011,Wheeler2013}. The sensor output was recorded, and converted into an airflow rate magnitude using the previously obtained airflow sensing characteristics as presented in Fig.~\ref{fig:ventilator} (a) and (b). sinusoidal signals corresponding to the reciprocating airflow at each set frequency was successfully captured, confirming the sensor’s ability to detect periodic airflow variations.

Next, the airflow tidal volume was determined by integrating the area under the airflow rate curve. The measurement yielded a reciprocating airflow volume of approximately 10 mL in both the positive and negative directions, demonstrating consistent and reliable quantitative readings. The agreement between both directions indicates the sensor's accuracy in measuring bidirectional airflow. The slight discrepancy from the set value can be attributed to the inherent error of the artificial ventilator, which is estimated to be around 10\%. Despite this minor deviation, the experiment successfully validated the sensor’s capability to provide accurate real-time airflow measurements in a controlled ventilator setting, showcasing its potential as an implant for respiration measurement.

\section{Conclusions}

In this study, we developed a stent-based flow sensor designed for real-time airflow measurement in the airways of laboratory animals. The stent was fabricated using biocompatible Ti, ensuring suitability for biomedical applications, and was processed using a combination of fiber laser digital processing and an origami-inspired folding technique. The integration of the sensing structure, fabricated through MEMS microfabrication technology, was achieved through a simple and effective insertion process, where the sensor film was secured at the center of the stent by utilizing its natural buckling mechanism. This approach allowed for seamless sensor placement, ensuring precise airflow detection without compromising the structural integrity of the stent. The fabricated stent implant was successfully expanded and installed in an airway-mimicking tube, demonstrating its stability and effectiveness in airflow sensing.  

A proof-of-concept trial using an artificial ventilator confirmed the capability of the proposed sensor system to monitor respiration in real time. The results validated the feasibility of integrating MEMS-based sensing technology with a metallic stent for preclinical respiratory assessments. This innovative approach has the potential to improve respiratory monitoring in animal studies, providing researchers with a more accurate and reliable method for evaluating airflow dynamics. Future work will focus on in vivo testing and further optimizing the sensor's sensitivity and durability for long-term implantation. Prior to such clinical translation, the sensor will be characterized at the target body temperature and pulmonary tract humidity under various expected breathing patterns of disease models. The influence of potential mucus accumulation and airway inflammation to the sensor response will also be elucidated. Wireless implementation such as using bluetooth connection, as well as power management using batteries or supercapacitors will be considered for real practical application. By advancing airflow measurement techniques in preclinical research, this study contributes to the development of more effective drug evaluations and a better understanding of respiratory diseases.

\section*{Acknowledgments}

This work was partially supported by the Japan Society for the Promotion of Science (JSPS) KAKENHI Grant-in-Aid for Early-Career Scientists under Grant 24K21094.












\newpage

\end{document}